\documentclass[12pt]{article}
\usepackage{bbm}
\usepackage{epsfig}
\usepackage{array}
\usepackage{float}
\usepackage{dsfont}
\usepackage{amstext}
\usepackage{cite}

\usepackage{a4}

\usepackage{a4wide}

\def\ra{\rightarrow}
\def\be{\begin{equation}}
\def\ee{\end{equation}}
\def\gs{\mathrel{
   \rlap{\raise 0.511ex \hbox{$>$}}{\lower 0.511ex \hbox{$\sim$}}}}
\def\ls{\mathrel{
   \rlap{\raise 0.511ex \hbox{$<$}}{\lower 0.511ex \hbox{$\sim$}}}}
\newcommand{\obb}{0\mbox{$\nu\beta\beta$}}
\newcommand{\onbb}{neutrino-less double beta decay}
\newcommand{\ba}{\begin{array}{c}}
\newcommand{\baz}{\begin{array}{cc}}
\newcommand{\bad}{\begin{array}{ccc}}
\newcommand{\bav}{\begin{array}{cccc}}
\newcommand{\baf}{\begin{array}{ccccc}}
\newcommand{\bea}{\begin{equation} \begin{array}{c}}
\newcommand{\eea}{ \end{array} \end{equation}}
\newcommand{\ea}{\end{array}}
\newcommand{\D}{\displaystyle}

\newcommand{\meff}{\mbox{$| m_{ee} |$}}
\newcommand{\imeff}{\mbox{$ \left| \frac{1}{M_{ee}} \right|$}}

\begin{document}

\title{
\vskip .4cm
\bf \Large 
Inverse Neutrino-less Double Beta Decay Revisited: Neutrinos, 
Higgs Triplets and a Muon Collider}
\author{\\  
Werner Rodejohann\thanks{email: 
\tt werner.rodejohann@mpi-hd.mpg.de}
\\ \\
{\normalsize \it Max--Planck--Institut f\"ur Kernphysik,}\\
{\normalsize \it  Postfach 103980, D--69029 Heidelberg, Germany} 
}
\date{}
\maketitle
\begin{abstract}
\noindent  
We revisit the process of inverse neutrino-less double
beta decay ($e^- e^- \ra W^- W^-$) at future linear colliders. The
cases of Majorana neutrino and Higgs triplet exchange are considered. 
We also discuss the processes $e^- \mu^- \ra W^- W^-$ and 
$\mu^- \mu^- \ra W^- W^-$, which are motivated by the possibility of
muon colliders. For heavy neutrino exchange and 
center-of-mass energies larger than 1 TeV, we show that masses up to
$10^6$ $(10^5)$ GeV could be probed for $ee$ and $e\mu$ machines,
respectively. The stringent limits for mixing of heavy neutrinos with
muons render $\mu^- \mu^- \ra W^- W^-$ less promising, even though
this process is not constrained by limits from neutrino-less double
beta decay. If Higgs triplets are responsible for inverse neutrino-less double
beta decay, observable signals are only possible if a very narrow 
resonance is met. We also consider unitarity aspects of the process in
case both Higgs triplets and neutrinos are exchanged. An exact see-saw
relation connecting low energy data with heavy neutrino and triplet 
parameters is found. 

\end{abstract}

\newpage

\section{\label{sec:intro}Introduction}
Observation of Lepton Number Violation (LNV) would show that 
neutrinos are Majorana particles \cite{SV} and would add most interesting 
information on the origin of neutrino masses. In particular, 
the process 
\be \label{eq:ee} 
e^- e^- \ra W^- W^- \, , 
\ee 
called ``inverse \onbb'', has frequently been 
proposed as a probe of LNV and new physics in general 
\cite{rizzo,belanger,other,IandIINU,triplet,triplet1}. 
Running a future linear collider in an $e^- e^-$ mode would allow looking 
for this and other lepton number violating and conserving 
processes \cite{ee}. 

We study inverse \onbb~here in the presence of Majorana neutrinos and 
a Higgs triplet. We also point out that the currently  
discussed muon colliders \cite{mc,geer} would allow to search for
lepton number (and flavor) violating processes like 
\be \label{eq:em}
\mu^- \mu^- \ra W^- W^- \, 
\mbox{ (and } e^- \mu^- \ra W^- W^-) \, , 
\ee
if the machines are run in a like-sign lepton mode. 
Physics potential of like-sign muon collisions has also been 
discussed in Refs.~\cite{mumu}, and is mentioned as a possibility 
in Refs.~\cite{mc}, where the prospects and technology of muon colliders are 
outlined (see \cite{rev} for a recent review on the current status of
muon collider research). 
We summarize the model-independent limits on heavy neutrino and Higgs 
triplet parameters which are relevant to these processes and give the 
corresponding values for the cross sections. 
We show in which situations the processes are observable. 
For heavy neutrino exchange, we show that in electron-electron collisions 
masses up to $10^6$ GeV could be probed, while like-sign $e\mu$
machines can reach $10^5$ GeV. The process $\mu^- \mu^- \ra W^- W^-$
is less promising, due to strong constraints on mixing of heavy neutrinos with
muons (note that this process is not constrained by limits from neutrino-less double
beta decay). If Higgs triplets are exchanged in inverse neutrino-less double
beta decay, observable signals are unlikely unless a very narrow 
resonance is met. 
We stress already here that we do not consider situations in which 
there are cancellations. This means we assume in the processes only
the exchange of one heavy neutrino or of one Higgs triplet, and not
the cases in which several neutrinos are present or both a triplet and
a heavy neutrino are present. However, we note that if  
both terms are present (as in the type I + II see-saw mechanism) there is an  
exact see-saw relation, which uses low energy data coming 
from neutrino-less double beta decay and other neutrino data to 
constrain a particular combination of high energy 
(i.e., heavy neutrino and Higgs triplet) parameters. It generalizes a 
previously discussed formula for the type I see-saw \cite{Xing_meff,ich}. 
We also comment 
on unitarity of the cross section $e^- e^- \ra W^- W^-$ in case of 
heavy neutrinos and triplets being simultaneously present. An exact
see-saw relation turns out to be helpful there.
 \\

The paper is build up as follows: in Section \ref{sec:0vbb} we discuss the 
present limits on parameters relevant for inverse neutrino-less double 
beta decay. Those come from neutrino-less double beta decay, 
global fits and other data, in particular lepton flavor violation. 
 Section \ref{sec:h} discusses the 
process of inverse \onbb~for heavy Majorana neutrinos, 
while Section \ref{sec:t} discusses 
the situation with a Higgs triplet. In Section \ref{sec:unitarity} we 
argue that unitarity of the cross section is automatically fulfilled in case 
a triplet and heavy neutrinos are present. 
We conclude in Section \ref{sec:concl}.

\begin{figure}[t]
\begin{center}
\hspace{-.63cm}
\includegraphics[width=17cm,height=7cm]{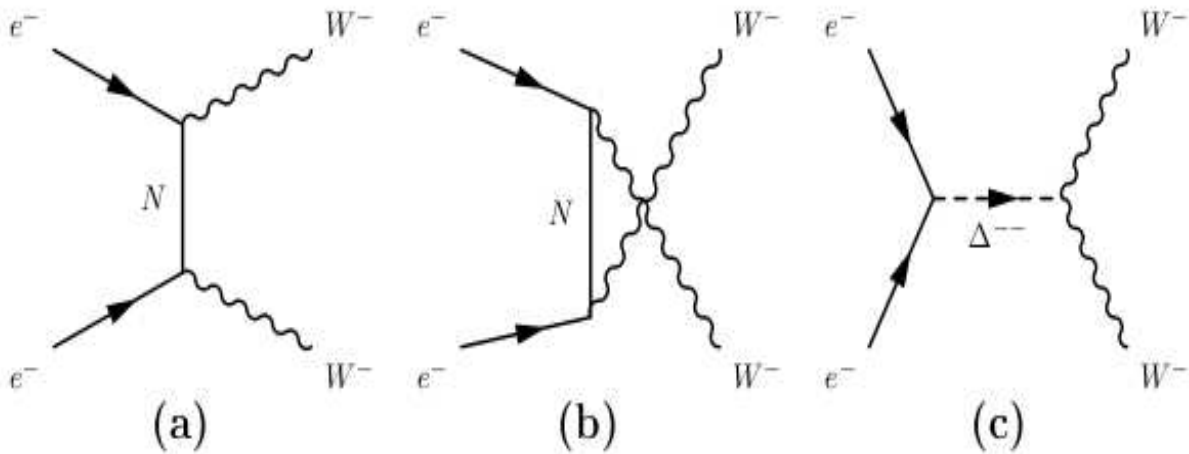} 
\caption{\label{fig:feynman}Diagrams for 
$e^- e^- \ra W^- W^-$ with Majorana neutrinos $N$ and a 
doubly charged Higgs scalar $\Delta^{--}$. Diagram 
(a) is the $t$-channel, (b) the 
$u$-channel and (c) the $s$-channel.}
\end{center}
\end{figure}

\section{\label{sec:0vbb}Neutrino-less Double Beta Decay, its 
Inverse, and Limits on Neutrino and Triplet Parameters}

Fig.~\ref{fig:feynman} shows the three 
different possible diagrams for inverse \onbb~in the presence of
Majorana neutrinos and a Higgs triplet. 
The diagrams (a) and (b) are connected 
to the diagram of \onbb~(\obb). This is the process 
$(A,Z) \ra (A,Z+2) + 2 e^-$, for which the electrons are 
outgoing and the $W^-$ couple to incoming $u$ and outgoing 
$d$ quarks. Indeed, as each vertex receives a factor $U_{ei}$ and the
propagator of the neutrino introduces a term $m_i/(q^2 - m_i^2)$, 
the dependence on neutrino parameters is the same as for \onbb.

Because a hypercharge $Y = 2$ Higgs triplet contains a doubly 
charged member ($\Delta^{--}$), diagram
(c) in Fig.~\ref{fig:feynman} is possible. The $\Delta^{--} $ 
can also lead to \obb~\cite{0vbbD}. 
One should note that the Higgs triplet contains also a singly charged 
scalar $\Delta^-$, which can contribute to \obb~as well 
(these are diagrams in which one $W^-$ is replaced by a 
$\Delta^-$). However, its  
coupling to quarks is suppressed by $v_L/v$, where $v_L$ is 
the vev of the neutral component of the triplet, and 
$v$ the vev of the SM doublet. Moreover, the triplets are presumably
heavier than the $W$. Hence, the diagrams for \obb~containing $\Delta^-$ are 
suppressed with respect to the diagram containing $\Delta^{--} $ and 
consequently there is a direct 
connection between inverse \onbb~and \onbb~also in 
scenarios with Higgs triplets.\\

We begin by studying constraints on light and heavy neutrino, 
as well as on Higgs triplet parameters from lepton number and
flavor violation. 
  
The most commonly assumed mechanism 
of \obb~is light neutrino exchange, for which the ``effective mass'' 
\meff~is constrained as follows \cite{APS}: 
\be \label{eq:light}
\meff \equiv \sum N_{ei}^2 \, (m_\nu)_i \ls 1 ~\rm eV \, . 
\ee
We have introduced here the notation that light neutrino masses 
are called $(m_\nu)_i$ and their mixing matrix is $N$. 
The above limit takes generously nuclear matrix element uncertainties 
into account. The most model-independent neutrino mass limit is 
2.3 eV from the Mainz and Troitsk experiments \cite{mainz}, and \meff~can 
not exceed this value. Hence, the above 
upper value of 1 eV is of the same order as the
``theoretical'' upper value of 2.3 eV, which is 
valid in case of quasi-degenerate 
light neutrinos (i.e., $(m_\nu)_1 \simeq (m_\nu)_2 
\simeq (m_\nu)_3 \equiv m_0$).

In case heavy neutrinos are exchanged in \onbb, the 
following quantity is constrained \cite{heavy} 
\be \label{eq:heavy1}
\D \imeff = \left| \sum S_{ei}^2 \, \frac{1}{M_i} \right| 
\ls 5 \cdot 10^{-8} \, {\rm GeV}^{-1} \, . 
\ee
Here heavy neutrino masses are called $M_i$ and the matrix describing
their mixing with leptons is called 
$S$. Note that at the current stage we have not discussed any see-saw mechanism 
connected to light and heavy neutrino masses, which would link
\meff~and \imeff.  
In what regards the mixing of electrons and muons with heavy neutral 
fermions, there are upper limits of \cite{nardi}
\be \label{eq:Smi}
\sum |S_{ei}|^2 \le 0.0052 ~ ,  ~~ \sum |S_{\mu i}|^2 \le 0.0001 \, , 
\ee
respectively, obtained from global fits, in particular of LEP data. 
Note that the limit on $|S_{\mu i}|^2$ is more stringent.  
Comparing with the \obb-limit in Eq.~(\ref{eq:heavy1}), the global 
one on $|S_{ei}|^2$ is stronger for masses $M_i \gs 10^5$ GeV.

The origin of the difference between 
\meff~and \imeff~is nothing but the two extreme limits of 
the fermion propagator of the Majorana neutrinos, which is 
central to the Feynman diagram of \obb: 
\be
\frac{{\slash \hspace{-.2cm}q} + m}{q^2 - m^2} 
\propto \left\{ 
\baz 
m & \mbox{ for } q^2 \gg m^2 \\
\frac{\D 1}{\D m} & \mbox{ for } q^2 \ll m^2
\ea
\right. \, .
\ee
Here $q$ denotes the momentum transfer in the process, 
which is around $100$ MeV and corresponds to $1/r$, where 
$r$ is the average distance of the 
two decaying nuclei. This helps us to understand roughly the numerical
value of the limit on \meff~and \imeff: the amplitude for 
light neutrino exchange is proportional to 
\be
{\cal A}_{\rm light} \simeq G_F^2 \, \frac{\meff}{q^2} \, ,  
\ee
while for heavy neutrinos it is proportional to 
\be
{\cal A}_{\rm heavy} \simeq G_F^2 \, \imeff \, . 
\ee
Therefore, a limit of 1 eV on \meff~corresponds to a limit on 
\imeff~of $10^{-7}$ GeV$^{-1}$. This rather crude estimate 
is surprisingly close the actual limit in Eq.~(\ref{eq:heavy1}), 
which takes the complicated nuclear physics into account. In the same 
approximation, we can estimate that the contribution of the doubly
charged Higgs triplet to \obb~has an amplitude proportional to 
\be
{\cal A}_{\rm triplet} \simeq G_F^2 \,  
\frac{h_{ee} \, v_L}{m_\Delta^2}\, ,   
\ee
where the factor $h_{ee}$ stems from the coupling of the triplet 
with the electrons, $v_L$ from the coupling of the triplet to the two $W$ 
and $1/m_\Delta^2$ is its  
propagator for $m_\Delta^2 \gg q^2$. Hence, we estimate 
the following limit on the triplet parameters from \obb:  
\be \label{eq:trip}
\left|\frac{h_{ee} \, v_L}{m_\Delta^2}\right| \ls 10^{-(7 \ldots 8)} 
\, {\rm GeV}^{-1} \, .
\ee
The triplet may be connected to neutrino mass because of the 
following term in the Lagrangian: 
\be 
{\cal L} = h_{\alpha \beta} \, 
\overline{L}_\alpha \, i \tau_2 \, 
\Delta \, L_\beta^c + h.c.
\ee
Here $h$ is a symmetric matrix, $\tau_2$ is a Pauli 
matrix, $L_\alpha$ a lepton doublet of flavor $\alpha = e, \mu, \tau$, and 
\be
\Delta = \left( \bad 
\Delta^+/\sqrt{2} & \Delta^{++} \\
\Delta^0 & -\Delta^+/\sqrt{2}
\ea
\right) 
\ee
contains the neutral, singly and doubly charged members of the Higgs
triplet. After the SM Higgs and the neutral component of the 
triplet obtain a vev 
($\langle \Phi \rangle = (0,v/\sqrt{2})^T$ and 
$\langle \Delta^0 \rangle = v_L/\sqrt{2}$) a direct 
contribution to the neutrino mass 
$m_L = \sqrt{2} \, v_L \, h$ arises. 
The electroweak $\rho$ parameter 
is modified to $\rho \simeq 1 - 2 \, v_L^2/v^2$, which leads to 
the constraint $v_L \ls 8$ GeV. 
Direct and model independent 
collider limits on the mass of the doubly charged triplet are 
$m_\Delta \gs 100$ GeV \cite{PDG}. 
It is interesting to compare this limit to limits stemming from
searches for lepton flavor violation (see e.g.~\cite{sugi,oldt}): 
\begin{eqnarray} \D \nonumber 
|h_{ee}|^2 \, |h_{e \mu}|^2 \left(\frac{250 \, \rm GeV}
{m_\Delta}\right)^4 & < &  2.1 \cdot 10^{-12} \,, \\ \D \nonumber 
|h_{ee}|^2 \, |h_{\tau \mu}|^2 \left(\frac{250 \, \rm GeV}
{m_\Delta}\right)^4 & < & 2.5 \cdot 10^{-7} \,, \\ \D \nonumber 
|h_{e\mu}|^2 \, |h_{\tau \mu}|^2 \left(\frac{250 \, \rm GeV}
{m_\Delta}\right)^4 & < & 1.3 \cdot 10^{-7} \, ,\\ \D 
|h_{\mu\mu}|^2 \, |h_{\tau \mu}|^2 \left(\frac{250 \, \rm GeV}
{m_\Delta}\right)^4 & < & 4.0 \cdot 10^{-7} \, ,\\ \D \nonumber 
|(h  h^\dagger)_{e \mu}| \left(\frac{250 \, \rm GeV}
{m_\Delta}\right)^4 & < & 6.5 \cdot 10^{-9} \, ,\\ \D \nonumber 
|(h  h^\dagger)_{e \tau}| \left(\frac{250 \, \rm GeV}
{m_\Delta}\right)^4 & < & 1.1 \cdot 10^{-4} \, ,\\ \D \nonumber 
|(h  h^\dagger)_{\mu\tau}| \left(\frac{250 \, \rm GeV}
{m_\Delta}\right)^4 & < &  1.4 \cdot 10^{-4} \, .\\ \D \nonumber 
\end{eqnarray}
Here we have used the current limits on the processes 
$\mu \ra 3  e$, $\tau \ra \mu \, 2  e$, 
$\tau \ra e \, 2 \mu$, $\tau \ra 3  \mu$, $\mu \ra e \, \gamma$, 
$\tau \ra e \, \gamma$ and $\tau \ra \mu \, \gamma$ \cite{PDG}. 
Constraints from $(g-2)_\mu$ (the anomalous magnetic moment of 
the muon, constraining $|h_{\mu\mu}|^2/m_\Delta^4$)
and muonium-antimuonium conversion 
(constraining $|h_{\mu\mu}|^2 \, |h_{ee}|^2/m_\Delta^4$) are very weak.

\section{\label{sec:h}Inverse Neutrino-less Double Beta Decay 
with Majorana Neutrinos}

\subsection{\label{sec:hee}$e^-e^-$ Collider}

For the process of inverse \obb~with Majorana neutrinos, diagrams 
(a) and (b) in Fig.~\ref{fig:feynman} apply. In the Appendix 
the lengthy formulae for the cross section including the mass of the 
$W$ can be found. In the useful and appropriate limit of negligible 
mass of the $W$ one has 
\be \label{eq:sigee}
\frac{ d \sigma}{d \cos \theta} = 
\frac{G_F^2}{32 \, \pi} \left\{ \sum m_i \, U_{ei}^2 \left( 
\frac{t}{t - m_i^2} + \frac{u}{u - m_i^2} 
\right) \right\}^2 \, .
\ee
Here $U_{\alpha i} = \left\{N_{\alpha 1}, N_{\alpha 2},
 N_{\alpha 3}, S_{\alpha 1},S_{\alpha 2},\ldots, S_{\alpha n}
\right\} $ is in our notation the general mixing matrix for 
the coupling of charged leptons with light and heavy neutrinos, whose
masses are given by 
$m_i = \{(m_\nu)_1, (m_\nu)_2, (m_\nu)_3, M_1, M_2,\ldots,M_n \}$. 
The extreme limits of the cross section are 
\be
\sigma(e^- e^- \ra W^- W^-) = \left\{ 
\baz \D 
\frac{G_F^2}{4 \, \pi} \, \left(U_{ei}^2 \, m_i\right)^2 & \mbox{for } 
s \gg m_i^2 \, , \\ \D 
\frac{G_F^2}{16 \, \pi}  s^2 
\left(\frac{U_{ei}^2}{m_i}\right)^2 & \mbox{for } 
s \ll m_i^2 \, .
\ea \right. 
\ee
We will comment below in Section \ref{sec:unitarity} 
on the apparent violation of unitarity in the 
limit of $s \ra \infty$. 
\begin{figure}[th]
\begin{center}
\includegraphics[width=8cm,height=7cm]{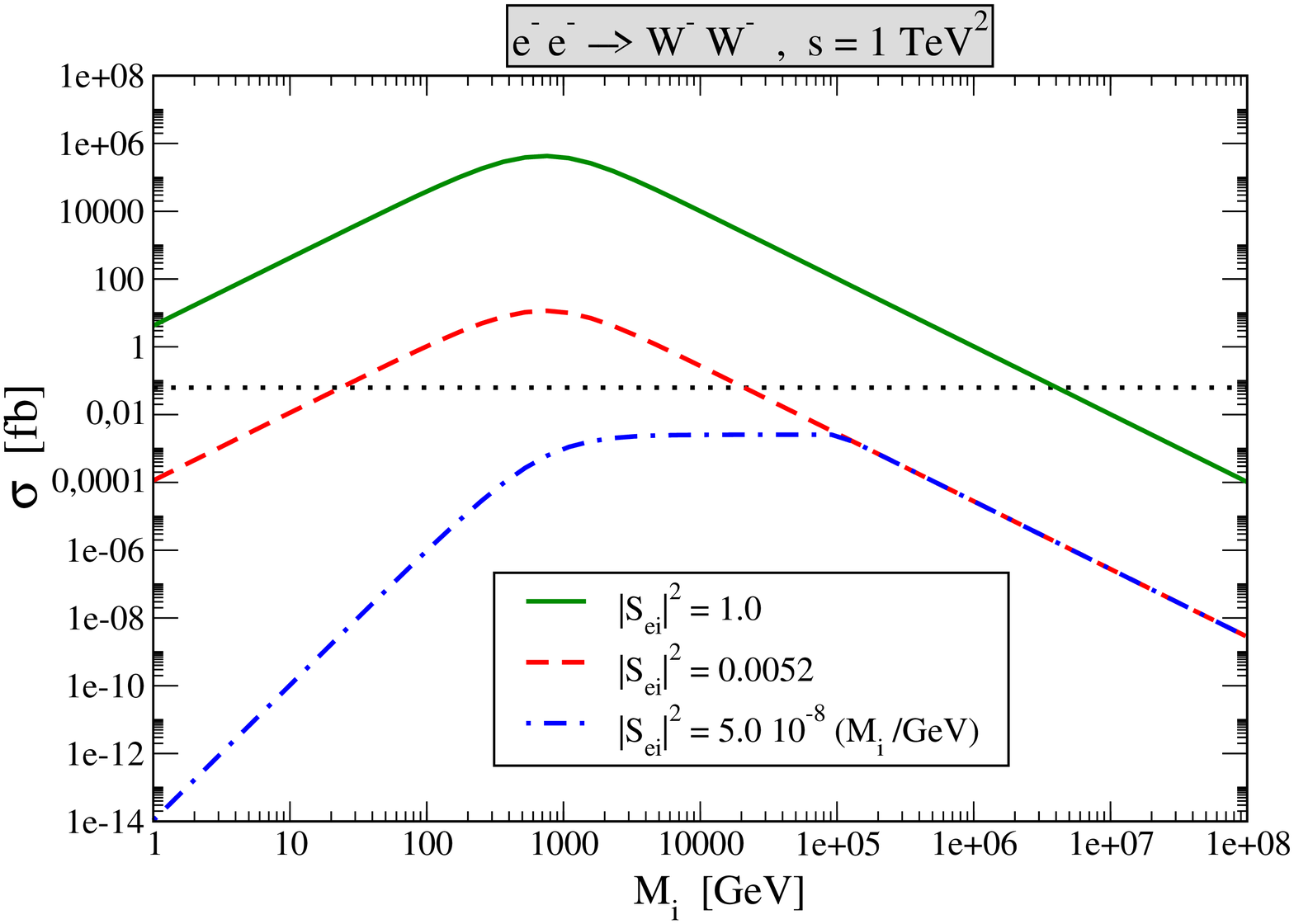} 
\includegraphics[width=8cm,height=7cm]{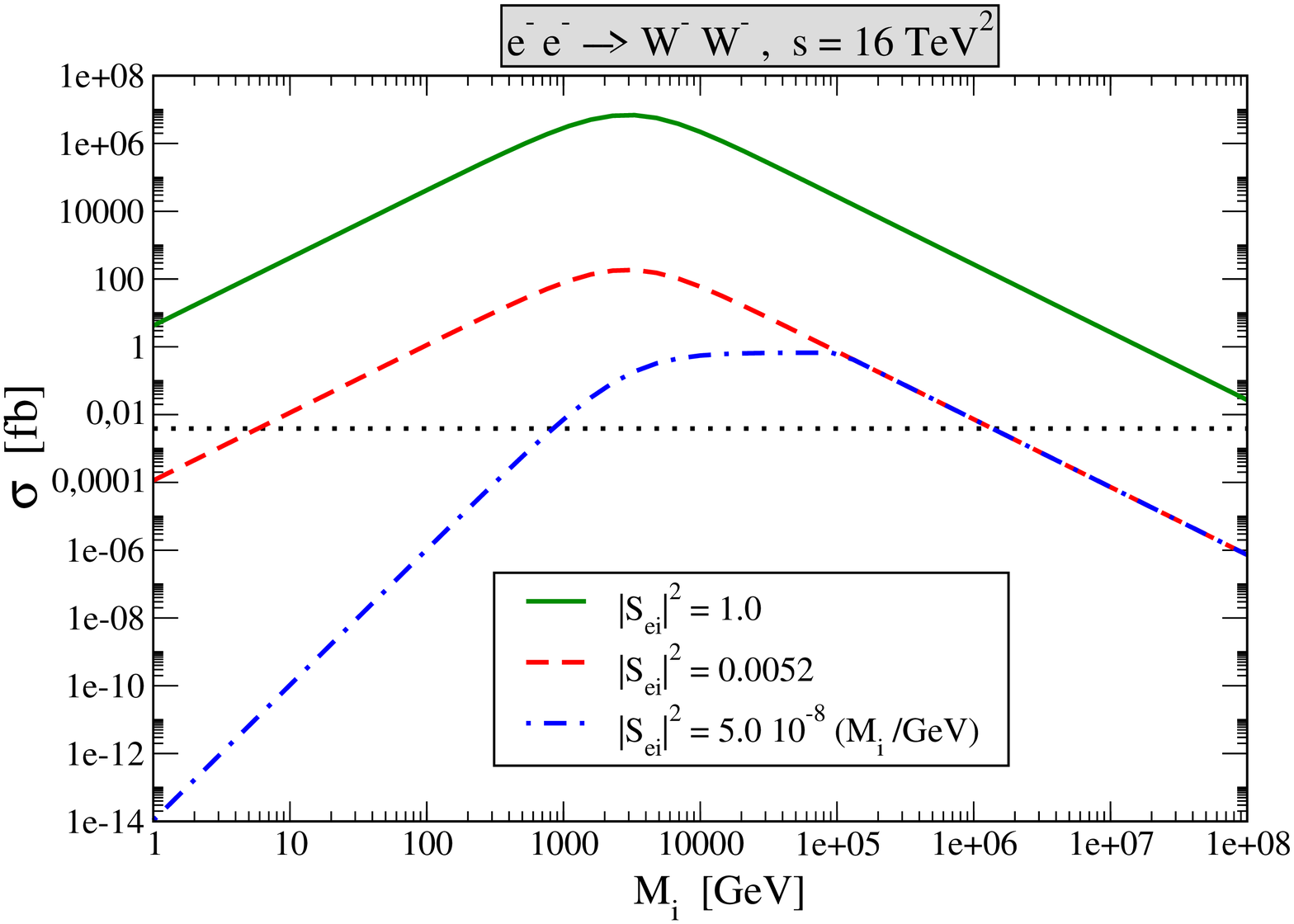}
\caption{\label{fig:ee}Cross section for $e^- e^- \ra W^- W^-$ with 
$\sqrt{s} = 1$ TeV (left) and $\sqrt{s} = 4$ TeV (right) and three  
limits for the mixing parameter $|S_{ei}|^2$. The dotted line 
corresponds to five events for an 
assumed luminosity of 80 $(s/{\rm TeV^2})$ fb$^{-1}$.}
\end{center}
\end{figure}
There are two interesting special cases for the cross 
section \cite{belanger}: 
\begin{itemize}
\item if only light active Majorana neutrinos contribute to 
the process, then we can bound the cross section as 
\be
\sigma(e^- e^- \ra W^- W^-) = \frac{G_F^2}{4 \, \pi} \, \meff^2 
\le 4.2 \cdot 10^{-18} \left(\frac{\meff}{1 \, {\rm eV} }
\right)^2 \, {\rm fb} \, . 
\ee 
\item if only heavy Majorana neutrinos contribute to 
the process, then we can bound the cross section 
using the \obb -limit from Eq.~(\ref{eq:heavy1}) as 
\be 
\sigma(e^- e^- \ra W^- W^-) = \frac{G_F^2}{16 \, \pi} \, s^2 
\, \imeff^2 \le 2.6 \cdot 10^{-3} \, \left(\frac{\sqrt{s}}{\rm TeV}
\right)^4 \left(\frac{\imeff}{5 \cdot 10^{-8} 
\, \rm GeV^{-1}} \right)^2 \, {\rm fb} \, . 
\ee 
\end{itemize}
Both numbers are far too small to be observable. In order to 
calculate the cross section for arbitrary neutrino masses, we have two
limits to take into account: first, the global limit on 
$|S_{ei}|^2$ from (\ref{eq:Smi}), and the limit on $S_{ei}^2/M_i$ from
\onbb~given in Eq.~(\ref{eq:heavy1}). Assuming the exchange of 
only one heavy neutrino results in Fig.~\ref{fig:ee}, where we plot 
the cross section for $e^- e^- \ra W^- W^-$ as a function of $M_i$ for
$\sqrt{s} = 1$ TeV and 
$\sqrt{s} = 4$ TeV. We give the curves for applying no limit, 
only the global one, and finally the \obb-limit in addition to the
global one. 
We indicate in the plot the cross section where five events for a 
luminosity of 80 $(s/{\rm TeV^2})$ fb$^{-1}$ \cite{belanger} 
would arise. From the plot one can see that the limit from
\obb~renders the process unobservable for $\sqrt{s} = 1$ TeV, while 
for $\sqrt{s} = 4$ TeV up to several $10^4$ events are possible. The
masses for which events are observable range from TeV to $10^3$ TeV. This
has to be compared with the situation at the LHC, where heavy Majorana
neutrinos are observable in the range 10 to 400 GeV for 100 fb$^{-1}$ 
\cite{han2} (see \cite{rev_LHC} for a review on neutrino production at
colliders).

\begin{figure}[th]
\begin{center}
\includegraphics[width=8cm,height=7cm]{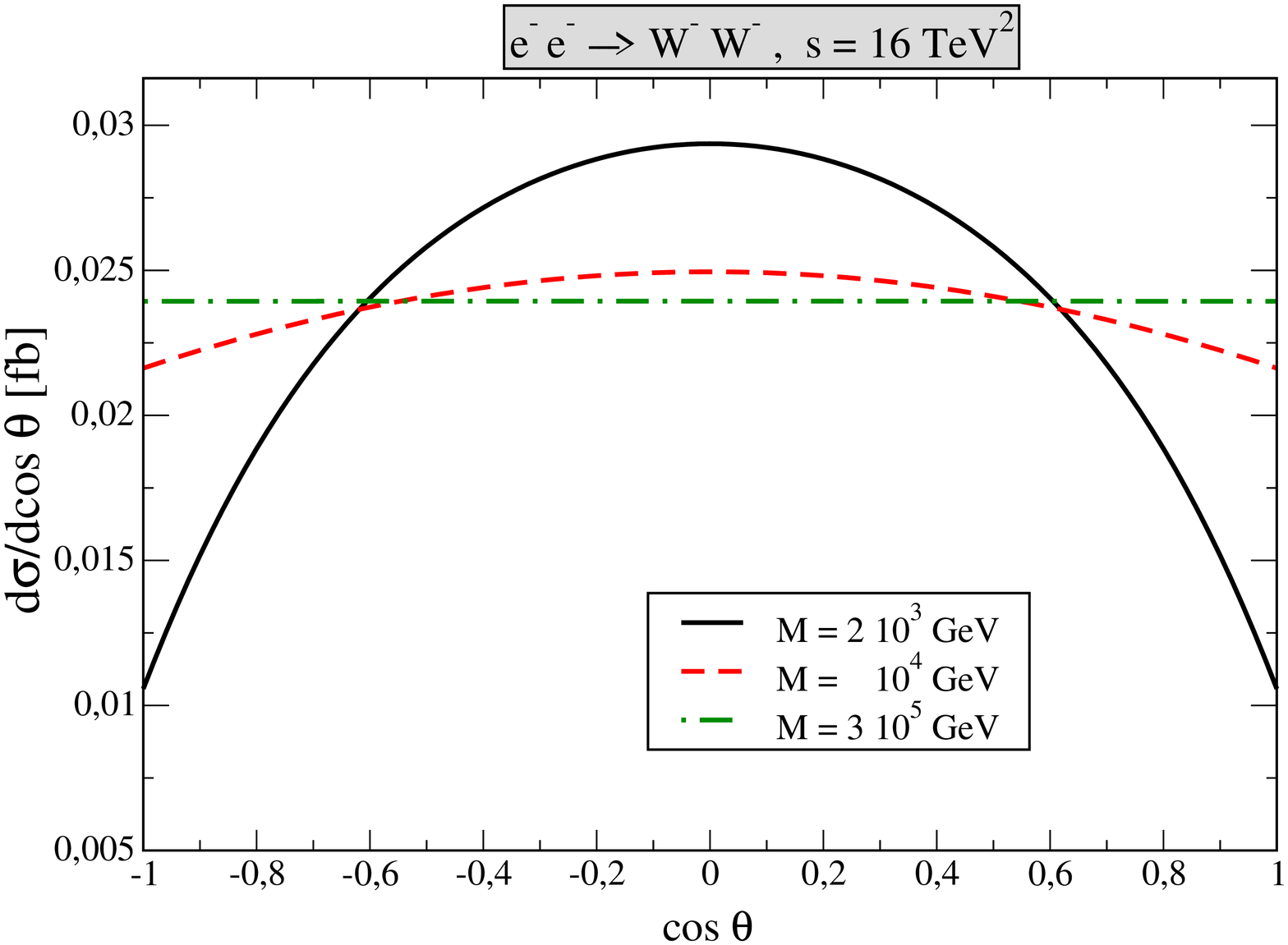} 
\caption{\label{fig:diff}
Differential cross section for $e^- e^- \ra W^- W^-$ with 
$\sqrt{s} = 4$ TeV and three different values of the neutrino mass,
with the mixing chosen such that the total cross sections are 
identical.} 
\end{center}
\end{figure}

In Fig.~\ref{fig:diff} we show the differential cross section 
$d\sigma/d\cos \theta$ for three different values of the neutrino mass
and the mixing $|U_{ei}|^2$ chosen such that the total cross sections 
are the same. 
We see that once $m_i \gg \sqrt{s}$ the differential 
cross section is essentially flat, which is also obvious from 
Eq.~(\ref{eq:sigee}).

\begin{figure}[th]
\begin{center}
\includegraphics[width=8cm,height=7cm]{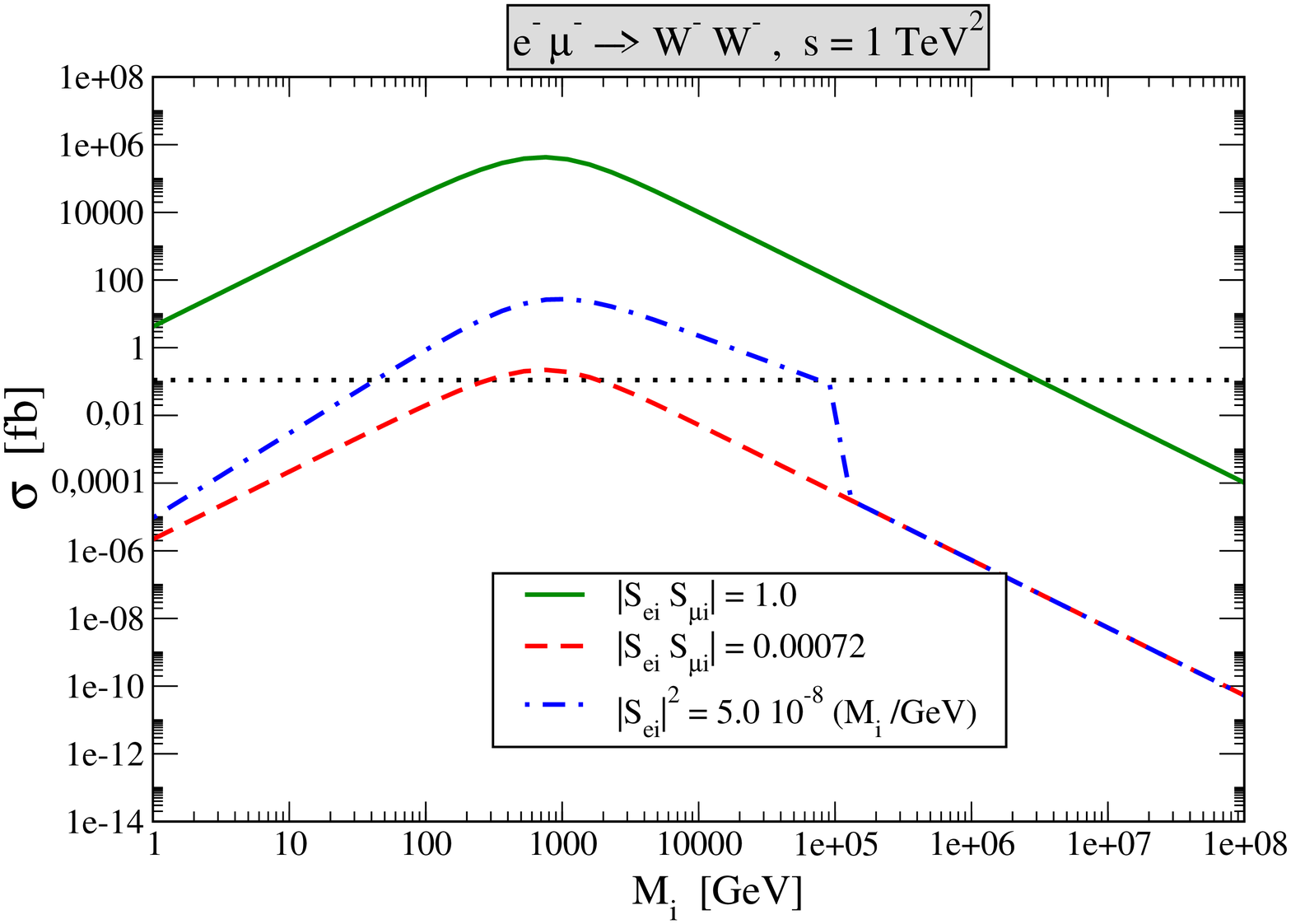} 
\includegraphics[width=8cm,height=7cm]{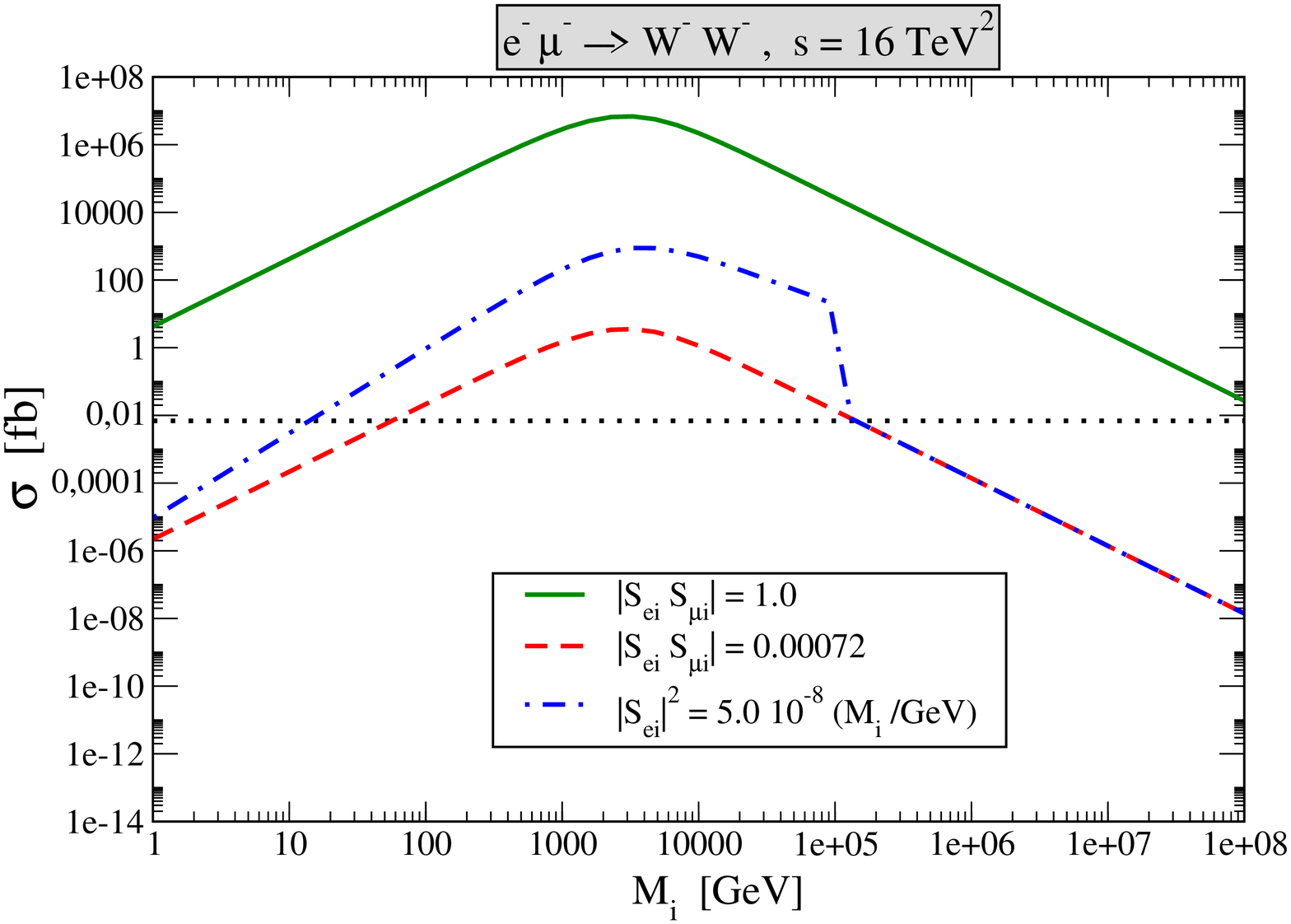}
\caption{\label{fig:em}Cross section for $e^- \mu^- \ra W^- W^-$ with 
$\sqrt{s} = 1$ TeV (left) and $\sqrt{s} = 4$ TeV (right) and three  
limits for the mixing parameter. The dotted line 
corresponds to five events for an 
assumed luminosity of 45 $(s/{\rm TeV^2})$ fb$^{-1}$.}
\end{center}
\end{figure}

\begin{figure}[th]
\begin{center}
\includegraphics[width=8cm,height=7cm]{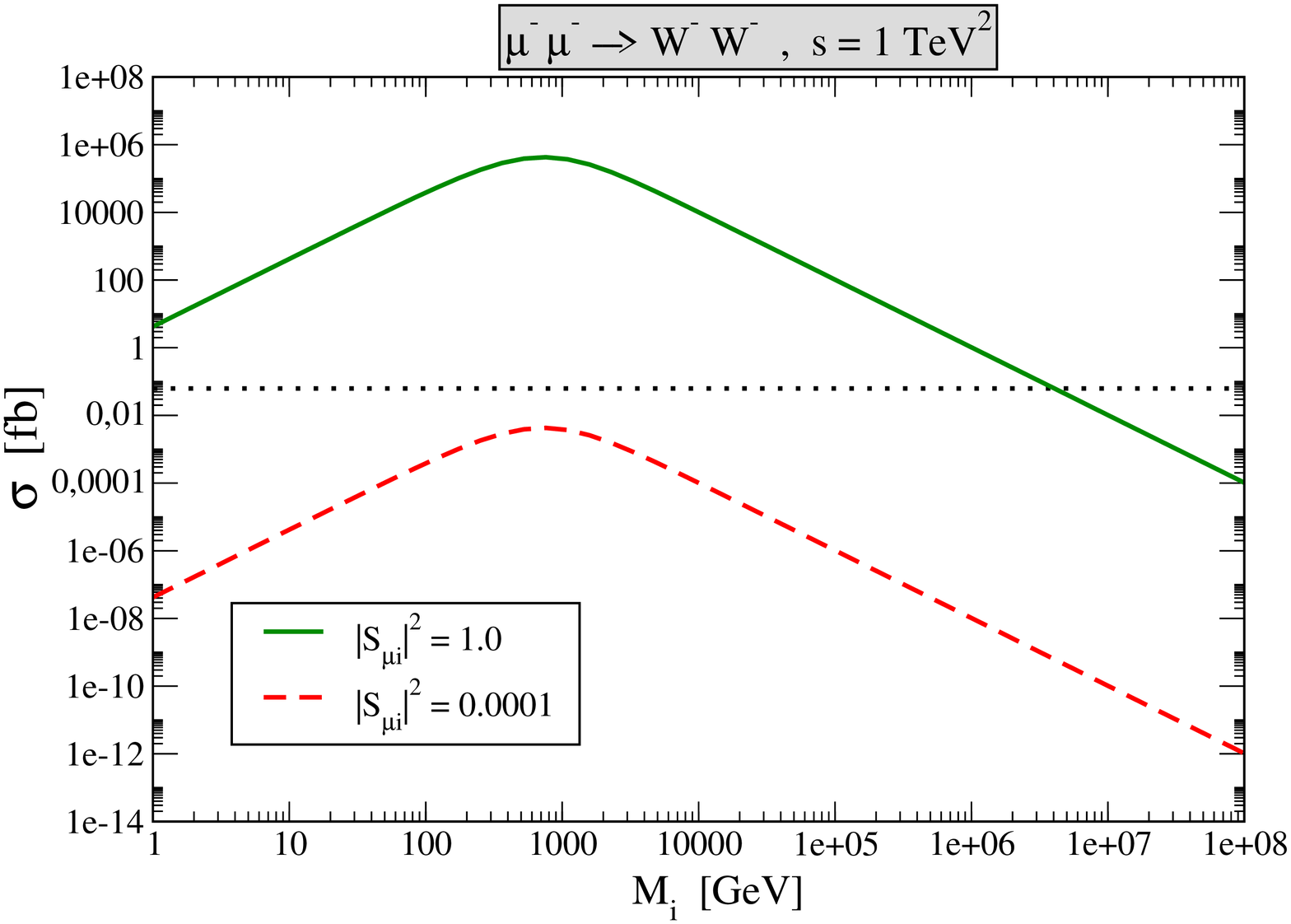} 
\includegraphics[width=8cm,height=7cm]{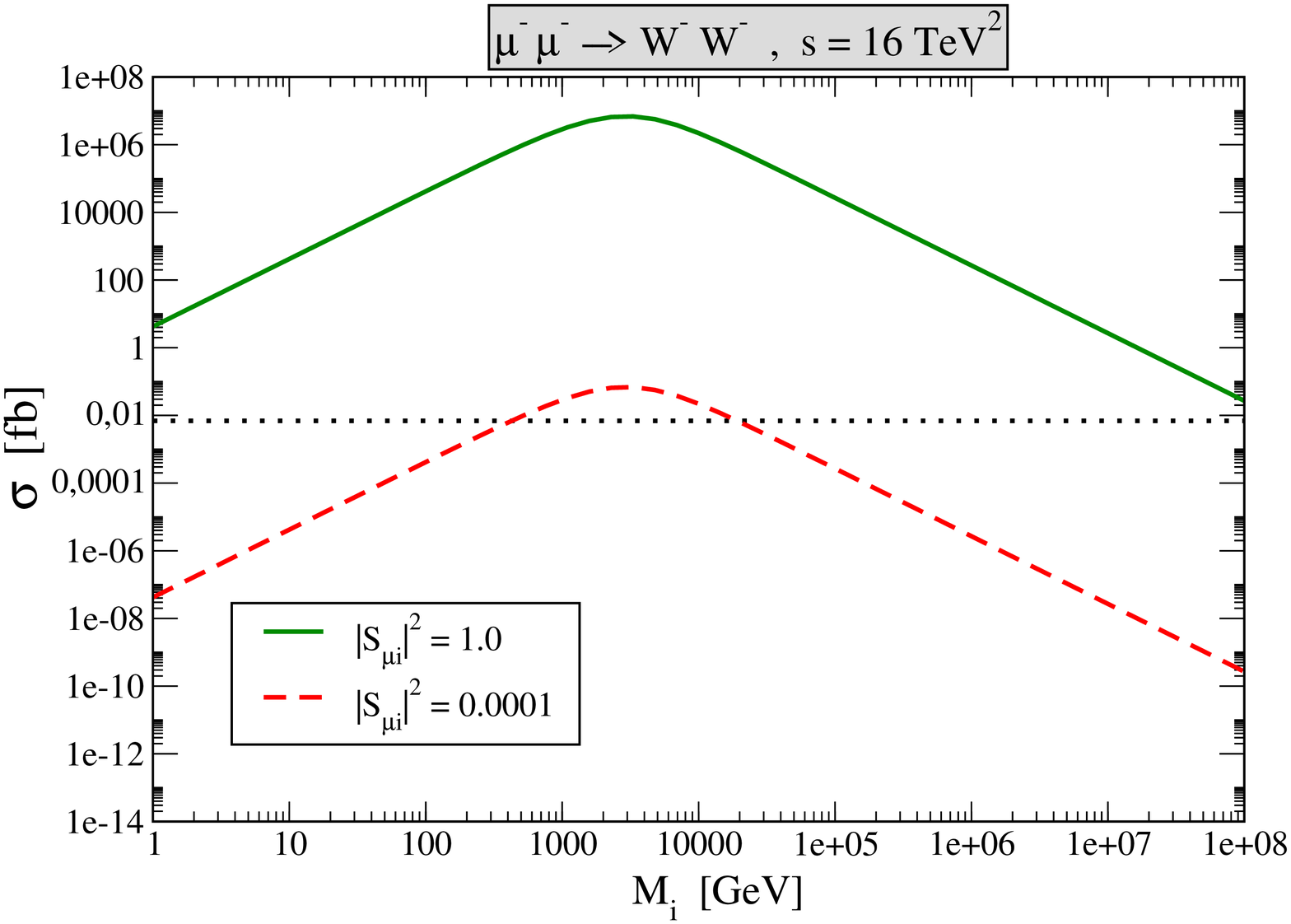}
\caption{\label{fig:mm}Cross section for $\mu^- \mu^- \ra W^- W^-$ with 
$\sqrt{s} = 1$ TeV (left) and $\sqrt{s} = 4$ TeV (right) and two  
limits for the mixing parameter $|S_{\mu i}|^2$. The dotted line 
corresponds to five events for an 
assumed luminosity of 45 $(s/{\rm TeV^2})$ fb$^{-1}$.}
\end{center}
\end{figure}

\subsection{\label{sec:hem}$e^-\mu^-$ Collider}
The plans of building a muon collider open up the possibility of 
studying the lepton number and flavor violating mode 
$e^- \, \mu^- \ra W^- \, W^-$. The differential cross section is 
\be
\frac{ d \sigma}{d \cos \theta} = 
\frac{G_F^2}{32 \, \pi} \left\{ \sum m_i \, U_{ei} \, U_{\mu i} \left( 
\frac{t}{t - m_i^2} + \frac{u}{u - m_i^2} 
\right) \right\}^2 \, .
\ee
If there are only light active neutrinos, then the 
cross section is proportional to $|m_{e\mu}|^2$, which is the 
$e\mu$ element of the mass matrix \cite{0vbb_ana,MR}. As this element can 
not be larger than 2.3 eV either, there is no hope of seeing the 
process in this case.  
In the case of heavy neutrinos contributing to $e^- \mu^- \ra W^-
W^-$, the limit on $\imeff$ influences this process as well.  
One needs to compare its effect with the global limit of 
$|S_{e i} \, S_{\mu i} | \ls 0.00072$. 
The cross section is given in Fig.~\ref{fig:em}. One can note that the
global limit can be stronger than the \obb -limit for large part of
the parameter space.

In what regards the luminosity of like-sign 
$e\mu$ or $\mu\mu$ machines, it is currently not clear what numbers 
can be achieved. Let us use the numbers of $\mu^+ \mu^-$ muon colliders as
examples. According to Ref.~\cite{mc}, 
integrated luminosities of 45 $(s/{\rm TeV^2})$ fb$^{-1}$, 
where we have assumed a year of $10^7$ s, are possible. We will 
take this value in the following for both $e\mu$ and $\mu\mu$ 
like-sign collisions. While large uncertainty is presumably 
associated with this
value, our results are easy to modify once more realistic estimates are
present. 

From the plots one can see that for $\sqrt{s} = 1$ TeV there is only a
tiny window around (400 - 600) GeV 
where a few events may happen, but for $\sqrt{s} = 4$ TeV 
up to a few 100 events between 100 and $10^5$ GeV 
are possible. The situation is thus slightly worse than 
for $ee$ collisions, even though there are strong constraints from 
neutrino-less double beta decay on $S_{ei}^2/M_i$. 
The reason is that the global limits on $|S_{\mu i}|^2$ are significantly 
stronger than on $|S_{ei}|^2$.

\subsection{\label{sec:hmm}$\mu^-\mu^-$ Collider}
Finally, let us discuss the possibility of a 
$\mu^- \mu^-$ mode. The cross section is  
\be
\frac{ d \sigma}{d \cos \theta} = 
\frac{G_F^2}{32 \, \pi} \left\{ \sum m_i \, U_{\mu i}^2 \left( 
\frac{t}{t - m_i^2} + \frac{u}{u - m_i^2} 
\right) \right\}^2 \, .
\ee
The only constraint comes from the global limit in 
Eq.~(\ref{eq:Smi}), which however is rather strong. 
Fig.~\ref{fig:mm} shows the cross section. 
From the plots one can see that for $\sqrt{s} = 4$ TeV there is only a 
smallish window between 400 and $10^4$ GeV in which up to a few 
10 events are possible. \\

We conclude from this Section that like-sign $ee$  
lepton collisions are most promising to search for heavy Majorana
neutrinos, and to constrain the parameter space of mixing matrix 
elements and mass. However, the center-of-mass energies should 
exceed 1 TeV. 
The already rather stringent limit on the mixing 
of heavy neutrinos with muons renders like-sign $e\mu$ collisions 
a bit less promising, and $\mu\mu$ facilities show little prospects to
determine LNV due to Majorana neutrinos.  
As a numerical example, for $M_i = 1.5$ TeV and $\sqrt{s} = 4$ TeV, 
a like-sign $e\mu$ or $ee$ collider would generate 5 events even for 
$|S|^2 = 3 \cdot 10^{-5}$, and improvement on the present 
bound of $|S|$ by one to two orders of magnitude would be possible 
(here $|S|^2$ denotes the respective combination of mixing parameters). 
For $M_i = 2 \cdot 10^5$ GeV, 5 events are possible even for 
$|S|^2 = 7 \cdot 10^{-4}$, resulting in an improvement of the bound on
the mixing by one order of magnitude.\\

Note that such limits and considerations apply most probably not 
for heavy neutrinos of the type I see-saw mechanism. In its natural 
form there is a clash between production of 
colliders and TeV scale masses of the heavy neutrinos: the 
mixing of the heavy neutrinos 
with the SM fermions is of order $|S| \sim m_D/M_R$, and the 
contribution to neutrino 
mass is $m_\nu \simeq m_D^2/M_R$. 
Since $m_\nu \ls$ eV, TeV-scale $M_R$ implies 
MeV-scale $m_D$, and hence $|S|$ is of order 
$10^{-6}$. 
However, the see-saw mechanism involves matrices, and 
highly fine-tuned scenarios in which the contributions 
of several heavy neutrinos compensate each other are possible, though
they seem rather unnatural and in particular unstable. We continue by studying 
inverse neutrino-less double beta decay in an often studied extension
of the Higgs sector.

\section{\label{sec:t}Inverse Neutrino-less Double Beta Decay 
with a Higgs Triplet}

The production of Higgs triplets in like-sign lepton collisions has been 
discussed also in Ref.~\cite{triplet,triplet1}. The cross section for 
$\alpha^- \beta^ - \ra W^- W^-$ is 
\bea \D 
\sigma = 2 \frac{d \sigma}{d \cos \theta} = 
\frac{G_F^2}{2 \pi} \, v_L^2 \, |h_{\alpha \beta}|^2 
\frac{(s - 2 \, m_W^2)^2 + 8 \, m_W^4}
{(s - m_\Delta^2)^2 + m_\Delta^2 \, \Gamma_\Delta^2} 
\, \sqrt{1 - 4 \, \frac{m_W^4}{s}} \\ \D 
\simeq 
\frac{G_F^2}{2 \pi} \, v_L^2 \, |h_{\alpha \beta}|^2 
\frac{s^2}{(s - m_\Delta^2)^2 + m_\Delta^2 \, \Gamma_\Delta^2}
\, , 
\eea
where $\Gamma_\Delta$ is the width of the
$\Delta^{--}$. 
We note that, in case only a
Higgs triplet contributes to neutrino mass, the process 
$\alpha^- \beta^ - \ra W^- W^-$ can not take place if the
entry $(m_\nu)_{\alpha\beta}$ vanishes. 
Recall that $v_L \, h = m_L/\sqrt{2}$, where 
$m_L$ is the triplet contribution to neutrino mass. 
Hence $|v_L \, h_{\alpha \beta}|$ can not exceed 1 eV, 
unless there are cancellations between the triplet and 
another contribution to neutrino mass, e.g., a type I see-saw
term. Neglecting this unnatural possibility, 
$v_L \, h$ can be at most $m_\nu/\sqrt{2}$, and the 
order of $\frac{G_F^2}{4 \pi} \, |(m_\nu)_{\alpha \beta}|^2$ is 
$10^{-18}$ fb for $m_\nu \simeq$ eV. 
It is therefore clear that the 
resonance needs to be met in order to see an observable signal. 
On resonance ($s = m_\Delta^2$) we find 
\be
\sigma^{\rm res} = \frac{G_F^2}{2\pi} \, v_L^2 \, |h_{\alpha \beta}|^2 \, 
\frac{m_\Delta^2}{\Gamma_\Delta^2} \, .  
\ee
Assuming 40 inverse femtobarn of luminosity and asking for more than 5 events
gives the requirement $m_\Delta/\Gamma_\Delta \gs 10^8$. 

We need to discuss the width of the triplet. 
Since the mass of the $\Delta^{--}$ is very close to
the mass of the $\Delta^-$ and $\Delta^0$, decays in final states
containing the other members of the triplet are very much
suppressed\footnote{The mass splitting due to electroweak corrections
between the doubly and singly charged members (if they have initially 
the same mass) is of order 
$G_F \, m_W^3$ \cite{minimal} and therefore too small to allow for
decays such as $\Delta^{--} \ra \Delta^- W^-$.}. 
The other decays of interest are into like-sign lepton pairs 
\be
\Gamma_\ell^{\alpha \beta} 
\equiv \Gamma(\Delta^{--} \ra \alpha^-  \beta^-) = 
\frac{|h_{\alpha \beta}|^2 }{4 \pi \, (1 + \delta_{\alpha \beta})}
m_\Delta \simeq 19.9 \, \frac{|h_{\alpha \beta}|^2 }{(1 +
\delta_{\alpha \beta})} \left(\frac{m_\Delta}{\rm 250 \, GeV} \right)
\, \rm GeV\, , 
\ee
and into a pair of $W$: 
\begin{eqnarray}
\Gamma_W \equiv \Gamma(\Delta^{--} \ra W^-  W^-) & = & \frac{v_L^2 \, g^4}{16 \pi \,
m_\Delta^2} \sqrt{m_\Delta^2 - 4 \, m_W^2} \left( 
2 + \frac{(m_\Delta^2 - 2 \, m_W^2)^2}{4 \, m_W^4}
\right) \\ \nonumber 
& \simeq & \frac{G_F^2}{2  \pi} \, v_L^2 \, m_\Delta^3 
\simeq 3.4 \cdot 10^{-10} 
\left( \frac{v_L}{\rm MeV} \right)^2
\left(\frac{m_\Delta}{\rm 250 \, GeV} \right)^3 \, \rm GeV \, .  
\end{eqnarray}
We have neglected the mass of the $W$ in the last row. 
Summing $\Gamma_\ell^{\alpha \beta}$ over all leptons and taking for
simplicity $h_{\alpha \beta} = h$ gives 
$m_\Delta/\sum \Gamma_\ell^{\alpha \beta} \simeq 1/|h|^2$. Thus,
for $|h| \simeq 10^{-4}$ the condition $m_\Delta/\Gamma_\Delta \simeq 10^8$
can be met. These order of magnitude estimates imply 
$v_L \simeq 10$ keV if $m_\nu \simeq 1$ eV. Indeed,
choosing for instance $m_\Delta = 500$ GeV, for such values of the
triplet vev the width in a $W$ pair is of order
$10^{-13}$ GeV, while $\sum \Gamma_\ell^{\alpha \beta} \simeq 10^{-6}$
GeV. We have therefore found a consistent scenario. 
Hence, the resonance condition is obtainable in cases in which
the decay into charged lepton pairs is favored. Interestingly, these
cases are the ones frequently studied in the literature \cite{lhc,han}. 
Pairs of $\Delta^{--}$ and $\Delta^{++}$ are produced mainly in
Drell-Yan processes, and the cross 
section \cite{han} between 250 and 800 GeV can
approximately be written as $\sigma \simeq 30 \, (250 \, {\rm
GeV}/m_\Delta)^4$ fb, so that 100 fb$^{-1}$ of luminosity can generate
a sizable amount of triplet pairs. This in turn would motivate the
study of $\alpha^- \beta^- \ra W^- W^-$ at a lepton collider, and to
scan the center-of-mass energy to make precision tests at
resonance. 

\begin{figure}[ht]
\begin{center}
\includegraphics[width=8cm,height=7cm]{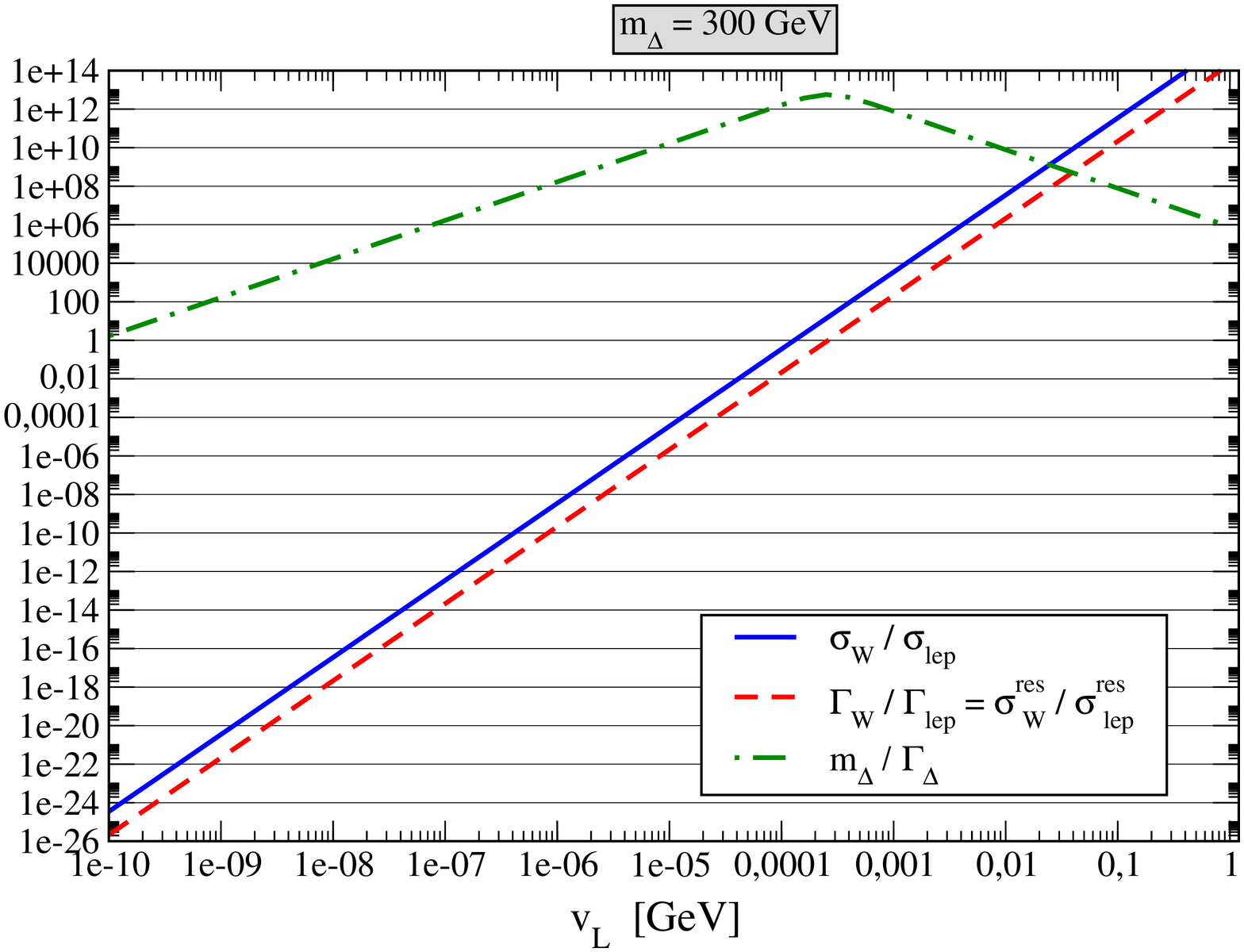} 
\includegraphics[width=8cm,height=7cm]{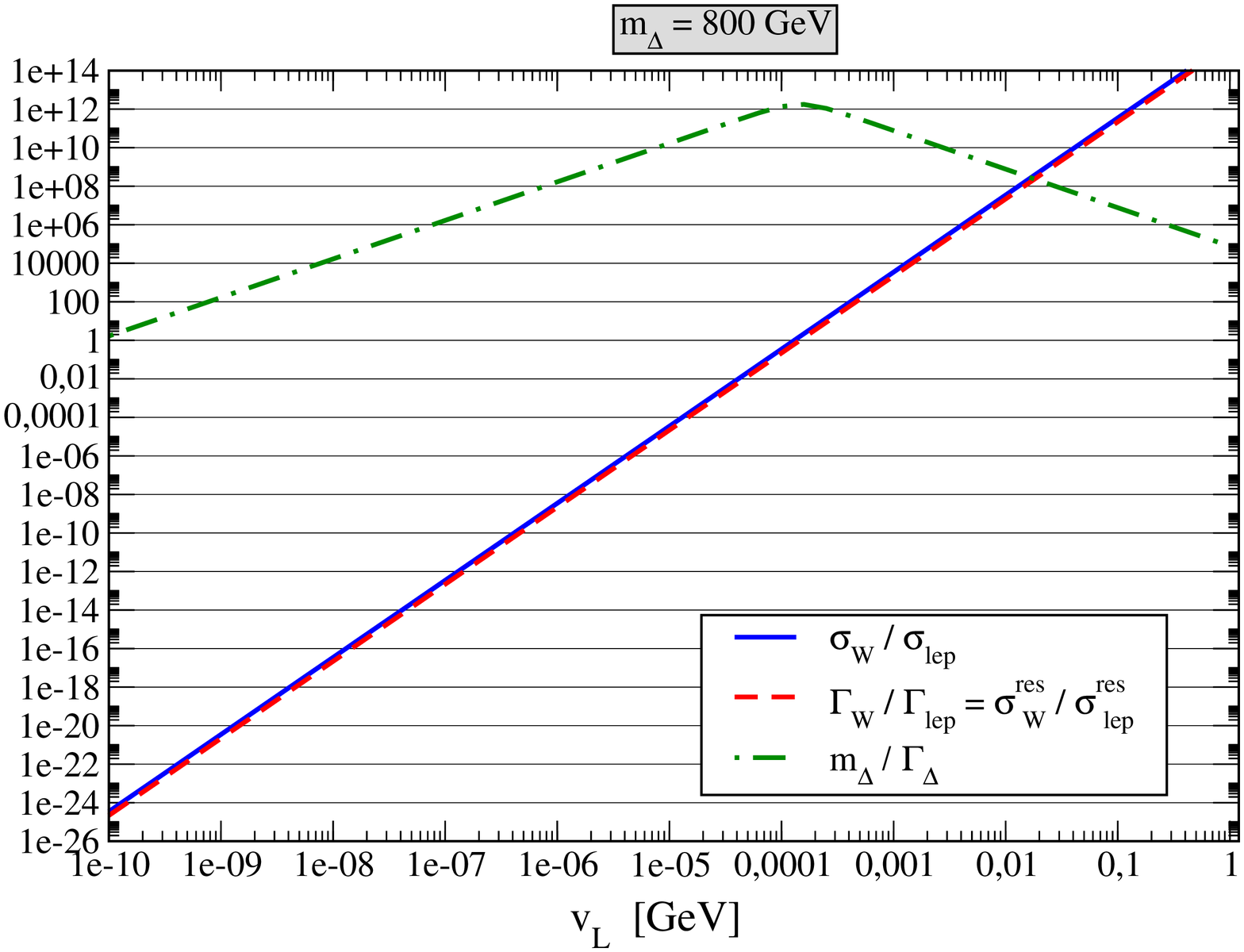} 
\caption{\label{fig:triplet2}The ratio of decay widths 
$\Gamma_W$ and $\Gamma_\ell$, as well as 
the ratio of cross sections (at $\sqrt{s} = 1$ TeV) as a function of
$v_L$. The ratio $m_\Delta$ to $\Gamma_{\Delta}$ is also plotted,
where $\Gamma_{\Delta}$ is the total width of the triplet. }
\end{center}
\end{figure}

One may wonder about another process in which a triplet is exchanged
in the $s$-channel, namely $\alpha^- \beta^- \ra \gamma^- \delta^-$, 
i.e., production of two like-sign leptons $\gamma$ and $\delta$ 
by collisions of two like-sign leptons $\alpha$ and $\beta$.  
The cross section is 
\be
\sigma(\alpha^- \beta^- \ra \gamma^- \delta^-) = 
\frac{|h_{\alpha\beta}|^2 \, |h_{\gamma \delta}|^2}{4 \pi \, 
(1 + \delta_{\gamma\delta})} \, 
\frac{s}{(s - m_\Delta^2)^2 + m_\Delta^2 \, \Gamma_\Delta^2} \, . 
\ee
The ratio of the cross sections is 
\be
\frac{\sigma(\alpha^- \beta^- \ra W^- W^-)}
{\sigma(\alpha^- \beta^- \ra \gamma^- \delta^-)} 
\equiv \frac{\sigma_{WW}}{\sigma_{\rm lep}}
\simeq 
2 \frac{G_F^2 \, v_L^2 \, s}{|h_{\gamma \delta}|^2 /(1 +
\delta_{\gamma\delta})} 
\stackrel{\rm res}{\ra} \frac{\Gamma_W}{\Gamma_\ell^{\gamma\delta}}
\, . 
\ee
At resonance, the ratio of cross sections equals the ratio of decay
widths. In Fig.~\ref{fig:triplet2}
we show for two values of $m_\Delta$ the ratio of decay widths 
$\Gamma_W$ and $\Gamma_\ell$, as well as 
the ratio of cross sections (at $\sqrt{s} = 1$ TeV) as a function of
$v_L$. The ratio $m_\Delta$ to $\Gamma_{\Delta}$ is also plotted,
where $\Gamma_{\Delta}$ is the total width of the triplet. 
We demanded the neutrino mass matrix $m_L = \sqrt{2} \,
v_L \, h$ to be of order 0.1 eV with $h_{\alpha \beta} = h $.

The simultaneous requirement of 
$\frac{\sigma(\alpha^- \beta^- \ra W^- W^-)}
{\sigma(\alpha^- \beta^- \ra \gamma^- \delta^-)} \gg 1$ and 
$m_\Delta/\Gamma_{\Delta} \gs 10^8$ implies a certain region in 
$m_\Delta$--$v_L$ space, see Fig.~\ref{fig:triplet2}. 
A typical point is $v_L = 0.002$ GeV, leading to $h \simeq 3
\cdot 10^{-8}$ for $m_\nu = 0.1$ eV. For such small couplings the 
limits from lepton flavor violation given above are obeyed. 

The width of the $\Delta^{--}$ is extremely small, much smaller than 
the beam spread, which has been estimated to be about 
$R = 10^{-2} \, \sqrt{s}$ for $ee$ colliders \cite{ee} and 
$R = 4 \cdot 10^{-4} \, \sqrt{s}$ for muon colliders \cite{mc}. For instance, 
if $m_\Delta = 600$ GeV, then for $\sqrt{s} = 600$ GeV the cross
section is 50.1 fb, while for $\sqrt{s} = 599.995$ GeV the cross
section is only $1.4 \cdot 10^{-10}$ fb. 
Picturing the spread as a box of width $R$ and
convoluting the cross section over this box \cite{triplet1} 
will smear out the resonance and give a $1/(R \, \Gamma)$ 
instead of a $1/\Gamma^2$ dependence of the cross section, thus
reducing the cross section by several orders of magnitude. For
instance, with $m_\Delta = \sqrt{s} = 600$ GeV and a spread $R$ the
result is $1.3/R  \cdot 10^{-6}$ pb. 
We conclude that observing triplet induced 
inverse \obb~at like-sign lepton colliders is very
unlikely. 

The situation is better for $\alpha^- \beta^- \ra \gamma^-
\delta^-$, where with $h \simeq 0.1$ one estimates the cross section far
away from resonance to be of order 
$\simeq h^4/(4\pi \, s) \simeq 3 ~({\rm TeV}/\sqrt{s})^2$ fb, which could
lead to sizable event numbers. With maximal Yukawa couplings of $4\pi$
the cross section would be 
$\sigma \simeq (4\pi)^3/s \simeq 0.8 ~({\rm GeV}/\Gamma_{\Delta})^2$ mb.
At resonance one has again with $h \simeq 0.1$ for the cross section 
$\sigma \simeq 3~({\rm TeV}/\sqrt{s})^2$ nb. 
We will study the prospects of this process elsewhere.

\section{\label{sec:unitarity}On Unitarity of $e^- \, e^- \ra W^- W^-$
and the Type I + II See-Saw Mechanism}
It is a useful exercise to consider the cross section of $e^- e^- \ra
W^- W^-$ in the presence of both the triplet and heavy neutrinos, and
study the unitarity behavior of the process. 

Towards this, consider scenarios with fermion singlets and Higgs triplets. 
Such a scenario is called type I + II see-saw, while the presence of
only a triplet may be called type II see-saw (sometimes denoted 
triplet see-saw). The presence of only fermion singlets is called 
type I see-saw. 
We can write down a coupling of lepton doublets with the triplet, 
a Yukawa mass term for the coupling of lepton doublets with fermion 
singlets, and a direct bare mass term for the singlets: 
\be 
{\cal L} = h_{\alpha \beta} \, 
\overline{L}_\alpha \, i \tau_2 \, 
\Delta \, L_\beta^c + \overline{L}_\alpha \, (Y_D)_{\alpha i} 
\, \Phi \, N_{R, j} + 
\frac 12 \, \overline{N_{R, i}^c} \, (M_R)_{ij} \, N_{R, j} + h.c.
\ee
Here $\Phi = (\phi^+,\phi^0)^T$ is the SM Higgs doublet and 
$M_R$ is a symmetric matrix. 
After the SM Higgs and the neutral component of the 
triplet obtain a vev,  
the complete mass term containing the Dirac and Majorana 
masses can be written as 
\bea \label{eq:Ltot}
{\cal L} = \frac 12 \, \overline{\nu_L} \,  m_L \, \nu_L^c + 
\overline{\nu_L} \, m_D \, N_R + \frac 12 \, 
\overline{N_R^c} \, M_R \, N_R + h.c.  \\ 
 = 
\frac 12 \, (\overline{\nu_L}, \overline{N_R^c}) \left( 
\baz  
m_L & m_D \\
m_D^T & M_R
\ea \right) 
\left(\ba \nu_L^c \\ N_R \ea \right) + h.c. 
\equiv  
\frac 12 \, (\overline{\nu_L}, \overline{N_R^c}) \, 
{\cal M} \, \left(\ba \nu_L^c \\ N_R \ea \right) + h.c. ,    
\eea 
with $m_D = Y_D \, v/\sqrt{2}$ and $m_L = \sqrt{2} \, v_L \, h$. There are 
in general six eigenvalues, 
\be
m^{\rm diag} = {\rm diag}(m_1, m_2, m_3, m_4, m_5, m_6) 
\, 
\ee
arising from diagonalizing the full $6\times6$ mass 
matrix ${\cal M}$ by a unitary $6\times6$ matrix 
\be \label{eq:Uss}
U = 
\left( \baz 
N & S \\
T & V 
\ea \right) \mbox{ with } 
{\cal M} = 
 U 
 \left( 
\baz 
m_\nu^{\rm diag} & 0 \\
0 & M_R^{\rm diag} 
\ea \right) U^T \, .
\ee
Here $m_\nu^{\rm diag} = {\rm diag}((m_\nu)_1, (m_\nu)_2, (m_\nu)_3)$
contains the light ``active'' neutrino masses and 
$M_R^{\rm diag} = {\rm diag}(M_1, M_2, M_3)$ the heavy ones. 
This difference between light and heavy neutrinos is valid 
if $m_L$ is much smaller than $m_D$ and 
$M_R$ is much bigger than $m_D$.  
The entries of $S$ and $T$ are in this case of order $m_D/M_R$, and hence 
one can obtain the expression 
\be 
N^\dagger 
\left( m_L - m_D \, M_{R}^{-1} \, m_D^T \right) N^\ast 
\simeq m_\nu^{\rm diag} \, .
\ee
Therefore, the mixing matrix in type I see-saw scenarios 
is strictly speaking not unitary, since 
$N N^\dagger = \mathbbm{1} - S S^\dagger\neq \mathbbm{1}$. 
The other set 
of heavy eigenvalues of ${\cal M}$ is obtained from 
$V^\dagger \, M_R \, V^\ast \simeq M_R^{\rm diag}$. 
We have illustrated the approximate nature of these expressions 
with the symbol $\simeq$, but for the usual magnitude of 
$m_L$, $m_D$ and $M_R$ the implied non-unitarity of $N$ is completely 
negligible. 
The matrix $S$ characterizes the mixing of the light 
neutrinos with the heavy ones:
\be 
\nu_\alpha = N_{\alpha i} \, \nu_i + S_{\alpha i} \, N_i \, ,
\ee
where $\nu_i$ ($N_i$) are the light (heavy) neutrinos with 
$i = 1,2,3$ and $\alpha = e, \mu, \tau$. 
The masses $(m_\nu)_i$ and $M_i$, and the associated mixing matrix
elements $N$ and $S$ can be constrained by \onbb, see Section 
\ref{sec:0vbb}.

Note that the 11-element of Eq.~(\ref{eq:Uss}) 
together with Eq.~(\ref{eq:Ltot}) reads 
\be \label{eq:exactII}
U \, m^{\rm diag} \,U^T = N \, m_\nu^{\rm diag} \, N^T + S \, M_R^{\rm diag} \, S^T 
=  m_L 
\, .
\ee 
We stress that this is an exact relation. It generalizes 
the relation 
$N \, m_\nu^{\rm diag} \, N^T + S \, M_R^{\rm diag} \, S^T 
= 0$, which is valid in absence of a triplet contribution 
to neutrino mass and which has been discussed in 
Ref.~\cite{Xing_meff} and further studied in \cite{ich}. 
The relation links, in type I + II see-saw scenarios, the measurable 
light neutrino parameters (the PMNS matrix $N$ and the light neutrino
masses) with the heavy Majorana neutrinos and the Higgs triplet
couplings and vev. 
In particular, Eq.~(\ref{eq:exactII}) implies 
for the effective mass that
\be
\meff 
= \left|(m_L)_{ee} - \sum S_{ei}^2 \, M_i\right| \, . 
\ee 
Consequently, if the triplet contribution to \onbb~is negligible with
respect to the light and heavy neutrino exchange,  
the experimental 
limits on \meff~apply directly to this combination
of parameters: 
\be \label{eq:xing}
\left|(m_L)_{ee} - \sum S_{ei}^2 \, M_i \right| 
= \left| \sqrt{2} \, h_{ee} \, v_L  - \sum S_{ei}^2 \, M_i \right| 
\ls 1~\rm eV \, . 
\ee 
In any case, as mentioned above, 
$N \, m_\nu^{\rm diag} \, N^T$ can not exceed 2.3 eV, and this is the 
formal maximal value of  $|(m_L)_{ee} - \sum S_{ei}^2 \, M_i|$.  
Obviously, in type I + II see-saw scenarios there is the interesting
potential of cancellations between terms involving neutrino and
triplet parameters. The individual limits on them can thus 
be evaded, and interesting phenomenology can arise. 
In this letter, however, we have discussed only the cases in which the 
triplets and neutrinos dominate in  $e^- e^- \ra W^- W^-$, 
$e^- \mu^- \ra W^- W^-$ and $\mu^- \mu^- \ra W^- W^-$, respectively and 
will treat the effect of cancellations elsewhere. 
However, an interesting aspect regarding unitarity of the cross
sections in case neutrinos and triplets contribute to 
inverse \onbb~is worth discussing: while the
full expression for the cross section is given in the Appendix, 
taking the high energy limit of $\sqrt{s} \ra \infty$ and 
setting $m_W$ to zero gives 
\begin{eqnarray} \D \nonumber  
\sigma & =&  \frac{G_F^2}{4 \pi} \left( 
(U_{ei}^2 \, m_i)^2 + 2 \, v_L^2 \, h_{ee}^2 
- 2 \sqrt{2}\, v_L \, h_{ee} \, U_{ei}^2 \, m_i
\right)  \\ \D 
& = &  \frac{G_F^2}{4 \pi} \left( 
(U_{ei}^2 \, m_i)^2 + (m_L)_{ee}^2  
- 2 \, (m_L)_{ee} \, U_{ei}^2 \, m_i
\right)  \\ \D \nonumber  
&=& \frac{G_F^2}{4 \pi} \left( (U_{ei}^2 \, m_i) - (m_L)_{ee}  \right)^2 
= 0 \, . 
\end{eqnarray} 
In the last line we have used the exact type I + II see-saw 
relation Eq.~(\ref{eq:exactII}). Thus, the cross section becomes
exactly zero in the high-energy limit. Recall that in case of no 
cancellation the cross section would be a constant, i.e., the
amplitude would grow with $\sqrt{s}$, thus violating unitarity. 
The exact see-saw relation cures this. 
This observation generalizes
the findings in \cite{rizzo,belanger}, in which it was shown that in case
of type I see-saw the cross section is $G_F^2/(4 \pi) \, (U_{ei}^2 \,
m_i)^2$ which is equal to zero in type I see-saw scenarios 
(see Eq.~(\ref{eq:exactII}) for $m_L = 0$)\footnote{
That the Higgs triplet restores unitarity of the process has been
noted also in \cite{rizzo,belanger}.}. 
Note that the requirement of vanishing $U_{ei}^2 \, m_i$ 
means that there can not be only one neutrino: there must be
necessarily two or more in order to make the cross section vanish in
the high energy limit. However, if a Higgs triplet is present then one neutrino
is enough.

\section{\label{sec:concl}Conclusions}
Future lepton colliders may be run in a like-sign lepton mode, thereby
probing lepton number violation. Here we have studied inverse
neutrino-less double beta decay, $\alpha^- \beta^- \ra W^- W^-$, with 
$(\alpha,\beta) = (e,e), (e,\mu)$ and $(\mu,\mu)$. We have discussed
two sources of lepton number violation, namely heavy Majorana
neutrinos and Higgs triplets. The former possibility is shown (for
$ee$ and $e\mu$ collisions and center-of-mass energies larger 
than 1 TeV) to be observable for masses up to $10^6$ GeV, which 
has to be compared with an LHC reach not exceeding 400 GeV. 
Triplet effects are unlikely to be seen,  
as a very narrow resonance has to be met. Surprisingly, even though
no limits from neutrino-less double beta decay apply, like-sign
muon colliders are a less promising option, because of strong
constraints on heavy neutrino mixing with muons.

\vspace{0.3cm}
\begin{center}
{\bf Acknowledgments}
\end{center}
This work was supported by the ERC under the Starting Grant 
MANITOP and by the Deutsche Forschungsgemeinschaft 
in the Transregio 27 ``Neutrinos and beyond -- weakly interacting 
particles in physics, astrophysics and cosmology''. 

\renewcommand{\theequation}{A\arabic{equation}}
\setcounter{equation}{0}
\renewcommand{\thetable}{A\arabic{table}}
\setcounter{table}{0}

\begin{appendix}
\section{Cross Section including $m_W$}
The three possibilities for $e^- (p_1) \, e^- (p_2) \ra 
W^- (k_1, \mu) \, W^- (k_2, \nu) $ are shown in Fig.\
\ref{fig:feynman}. Here $p_{1,2}$ and $k_{1,2}$ are the momenta of the
particles and $\mu, \nu$ the Lorentz indices of the $W$ polarization
vectors. 
The matrix element is 
\be 
-i {\cal M} = -i \left({\cal M}_t + {\cal M}_u + {\cal M}_s \right) , 
\ee
where the subscript denotes whether it is the $t$, $u$ or $s$
channel. The vertex for $\Delta \, W \, W$ is $i \, \sqrt{2}  \, g^2 \, 
v_L \, g_{\mu \nu}$. In order to evaluate the cross section 
\be
\frac{d \sigma}{d \Omega} = \frac 12 \, \frac{1}{64 \pi^2 \, s} 
   \frac 14 |\overline{\cal M} |^2 \, 
\sqrt{\frac{\lambda(s,m_W^2,m_W^2)}{\lambda(s,0,0)}} \, ,
\ee
where the first $\frac 12$ is due to two identical particles in the
final state and $\lambda(a,b,c) = a^2 + b^2 + c^2 - 2 \,(a\,b + a\,c +
b\,c)$, we need  
\be
|\overline{\cal M} |^2 = 
|\overline{{\cal M}_t} |^2 + |\overline{{\cal M}_u} |^2 + 
|\overline{{\cal M}_s} |^2 + 2 \, {\rm Re} \left( 
\overline{ {\cal M}_t^\ast \, {\cal M}_u} + 
\overline{ {\cal M}_t^\ast \,  {\cal M}_s} + 
\overline{ {\cal M}_u^\ast \,  {\cal M}_s}
\right) . 
\ee
The result is 
\begin{eqnarray} \nonumber
|\overline{{\cal M}_t} |^2 & = &  \frac{g^4}{4 \, m_W^4} 
\, U_{ei}^2 \, U_{ej}^2 \, m_i \, m_j \left[ 
\frac{4 \, m_W^6 - 4 \, m_W^4 \, (t + u) - t^2 \, (t + u) 
+ 2 \, m_W^2 \, (t  + 2 \, u)}{(t - m_i^2) \,  (t - m_j^2)}
\right]\, , \\ \nonumber
|\overline{{\cal M}_u} |^2 & = &  |\overline{\cal M}_t |^2 \, (t \leftrightarrow
u) \, , \\ \nonumber
\overline{ {\cal M}_t^\ast \, {\cal M}_u} & = & 
\frac{g^4}{4\, m_W^4} \, U_{ei}^2 \, U_{ej}^2 \,
m_i \, m_j \left[ 
\frac{4 \, m_W^6 - 2 \, m_W^2 \, t \, u - t \, u \, (t + u)}
{(u - m_i^2) \,  (t - m_j^2)}
\right]\, , \\ \nonumber
|\overline{{\cal M}_s} |^2 & = &
2 \, \frac{g^4}{m_W^4}  \, v_L^2 \, h_{ee}^2 \, 
\frac{s \, (8 \, m_W^4 + (s - 2 \, m_W^2)^2)}{(s -
m_\Delta^2)^2} \, , \\ \nonumber
\overline{ {\cal M}_t^\ast \, {\cal M}_s} & = & 
\sqrt{2} \, \frac{g^4}{m_W^4}  \, v_L \, h_{ee} \, U_{ei}^2 \, m_i \, 
\frac{(2 \, m_W^2 - t - u) \, (4 \, m_W^4 + t \, (t + u))}
{(t - m_i^2) \, (s - m_\Delta^2)} \, , \\ \nonumber
\overline{ {\cal M}_u^\ast \, {\cal M}_s} & = & \overline{ {\cal
M}_t^\ast \, {\cal M}_s} \, (t \leftrightarrow u) \, .
\end{eqnarray}

\end{appendix}


\begin{thebibliography}{99}

\bibitem{SV}
J.~Schechter and J.~W.~F.~Valle,
  Phys.\ Rev.\  D {\bf 25}, 2951 (1982).



\bibitem{triplet}
 T.~G.~Rizzo,
  Phys.\ Rev.\  D {\bf 25}, 1355 (1982)
  [Addendum-ibid.\  D {\bf 27}, 657 (1983)];
 P.~H.~Frampton and D.~Ng,
  Phys.\ Rev.\  D {\bf 45}, 4240 (1992); 
 J.~Gluza and M.~Zralek,
  Phys.\ Rev.\  D {\bf 52}, 6238 (1995)
  [arXiv:hep-ph/9502284]; 
J.~Gluza,
  Phys.\ Lett.\  B {\bf 403}, 304 (1997)
  [arXiv:hep-ph/9704202]; 
 J.~F.~Gunion,
  Int.\ J.\ Mod.\ Phys.\  A {\bf 11}, 1551 (1996)
  [arXiv:hep-ph/9510350];
M.~Raidal,
  Phys.\ Rev.\  D {\bf 57}, 2013 (1998)
  [arXiv:hep-ph/9706279];
B.~Mukhopadhyaya and S.~K.~Rai,
  Phys.\ Lett.\  B {\bf 633}, 519 (2006)
  [arXiv:hep-ph/0508290].

\bibitem{triplet1}
F.~Cuypers and M.~Raidal,
  Nucl.\ Phys.\  B {\bf 501}, 3 (1997)
  [arXiv:hep-ph/9704224].


\bibitem{IandIINU}
P.~Duka, J.~Gluza and M.~Zralek,
  Phys.\ Rev.\  D {\bf 58}, 053009 (1998)
  [arXiv:hep-ph/9804372].





\bibitem{rizzo}T.~G.~Rizzo,
  Phys.\ Lett.\  B {\bf 116}, 23 (1982).

\bibitem{belanger}
G.~Belanger, F.~Boudjema, D.~London and H.~Nadeau,
  Phys.\ Rev.\  D {\bf 53}, 6292 (1996)
  [arXiv:hep-ph/9508317].



\bibitem{other}
D.~London, G.~Belanger and J.~N.~Ng,
  Phys.\ Lett.\  B {\bf 188}, 155 (1987);
V.~D.~Barger, J.~F.~Beacom, K.~M.~Cheung and T.~Han,
  Phys.\ Rev.\  D {\bf 50}, 6704 (1994)
  [arXiv:hep-ph/9404335]; 
J.~Gluza and M.~Zralek,
  Phys.\ Lett.\  B {\bf 362}, 148 (1995)
  [arXiv:hep-ph/9507269]; 
  Phys.\ Lett.\  B {\bf 372}, 259 (1996)
  [arXiv:hep-ph/9510407];
C.~A.~Heusch and P.~Minkowski,
  arXiv:hep-ph/9611353.

\bibitem{ee}J.~Brau {\it et al.}  [ILC Collaboration],
  arXiv:0712.1950 [physics.acc-ph]; 
NLC Design Group, 
Collider, Vol.~2, SLAC-R-0474-VOL-2 VOL-2 
{\tt http://www.slac.stanford.edu/accel/nlc/zdr}


\bibitem{mc} J.~C.~Gallardo {\it et al.},
{\it In the Proceedings of 1996 DPF / DPB Summer Study on New Directions for High-Energy Physics (Snowmass 96), Snowmass, Colorado, 25 Jun - 12
Jul 1996, pp R4}; 
 M.~M.~Alsharoa {\it et al.}  [Muon Collider/Neutrino Factory
                  Collaboration],
  Phys.\ Rev.\ ST Accel.\ Beams {\bf 6}, 081001 (2003)
  [arXiv:hep-ex/0207031]. 
Slides from a recent workshop on the topic can be 
found at {\tt 
http://www.fnal.gov/directorate/Longrange/Steering$\_$Public/ \newline
workshop-muoncollider.html}

\bibitem{geer}S.~Geer,
  Ann.\ Rev.\ Nucl.\ Part.\ Sci.\  {\bf 59}, 347 (2009).


\bibitem{mumu}
C.~A.~Heusch and F.~Cuypers,
  AIP Conf.\ Proc.\  {\bf 352}, 219 (1996)
  [arXiv:hep-ph/9508230]; 
M.~Raidal,
  Phys.\ Rev.\  D {\bf 57}, 2013 (1998)
  [arXiv:hep-ph/9706279].

\bibitem{rev}V.~Shiltsev,
  arXiv:1003.3051 [physics.acc-ph].



\bibitem{Xing_meff}
 Z.~Z.~Xing,
   Phys.\ Lett.\  B {\bf 679}, 255 (2009)
  [arXiv:0907.3014 [hep-ph]].



\bibitem{ich}
W.~Rodejohann,
  Phys.\ Lett.\  B {\bf 684}, 40 (2010)
  [arXiv:0912.3388 [hep-ph]].


\bibitem{0vbbD} R.~N.~Mohapatra and J.~D.~Vergados,
  Phys.\ Rev.\ Lett.\  {\bf 47}, 1713 (1981);
 L.~Wolfenstein,
  Phys.\ Rev.\  D {\bf 26}, 2507 (1982); 
W.~C.~Haxton, S.~P.~Rosen and G.~J.~Stephenson,
  Phys.\ Rev.\  D {\bf 26}, 1805 (1982).




\bibitem{APS}For a list of many experimental limits, 
F.~T.~.~Avignone, S.~R.~Elliott and J.~Engel,
  Rev.\ Mod.\ Phys.\  {\bf 80}, 481 (2008)
  [arXiv:0708.1033 [nucl-ex]]; 
 O.~Cremonesi,
  arXiv:1002.1437 [hep-ex].




\bibitem{mainz}
C.~Kraus {\it et al.},
  Eur.\ Phys.\ J.\  C {\bf 40}, 447 (2005)
  [arXiv:hep-ex/0412056]; 
  V.~M.~Lobashev,
  Nucl.\ Phys.\  A {\bf 719}, 153 (2003).


\bibitem{heavy}H.~V.~Klapdor-Kleingrothaus and H.~P\"as,
  arXiv:hep-ph/0002109; 
P.~Benes, A.~Faessler, F.~Simkovic and S.~Kovalenko,
  Phys.\ Rev.\  D {\bf 71}, 077901 (2005)
  [arXiv:hep-ph/0501295].



\bibitem{nardi}
 F.~M.~L.~Almeida, Y.~D.~A.~Coutinho, J.~A.~Martins Simoes and M.~A.~B.~do Vale,
  Phys.\ Rev.\  D {\bf 62}, 075004 (2000)
  [arXiv:hep-ph/0002024]; 
 E.~Nardi, E.~Roulet and D.~Tommasini, 
  Phys.\ Lett.\  B {\bf 344}, 225 (1995)
  [arXiv:hep-ph/9409310].





\bibitem{PDG} C.~Amsler {\it et al.}  [Particle Data Group],
  Phys.\ Lett.\  B {\bf 667}, 1 (2008).



\bibitem{sugi} A.~G.~Akeroyd, M.~Aoki and H.~Sugiyama,
  Phys.\ Rev.\  D {\bf 79}, 113010 (2009)
  [arXiv:0904.3640 [hep-ph]].


\bibitem{oldt}F.~Cuypers and S.~Davidson,
  Eur.\ Phys.\ J.\  C {\bf 2}, 503 (1998)
  [arXiv:hep-ph/9609487]; 
P.~H.~Frampton,
  Int.\ J.\ Mod.\ Phys.\  A {\bf 13}, 2345 (1998)
  [arXiv:hep-ph/9711281].



\bibitem{han2}T.~Han and B.~Zhang,
  Phys.\ Rev.\ Lett.\  {\bf 97}, 171804 (2006)
  [arXiv:hep-ph/0604064].



\bibitem{rev_LHC} F.~del Aguila, J.~de Blas, A.~Carmona and J.~Santiago,
  arXiv:1003.5799 [hep-ph].


\bibitem{MR}A.~Merle and W.~Rodejohann,
  Phys.\ Rev.\  D {\bf 73}, 073012 (2006)
  [arXiv:hep-ph/0603111].


\bibitem{0vbb_ana}
W.~Rodejohann,
  J.\ Phys.\ G {\bf 28}, 1477 (2002). 









\bibitem{lhc}E.~J.~Chun, K.~Y.~Lee and S.~C.~Park,
  Phys.\ Lett.\  B {\bf 566}, 142 (2003)
  [arXiv:hep-ph/0304069]; 
  A.~G.~Akeroyd and M.~Aoki,
  Phys.\ Rev.\  D {\bf 72}, 035011 (2005)
  [arXiv:hep-ph/0506176]; 
  A.~G.~Akeroyd, M.~Aoki and H.~Sugiyama,
  Phys.\ Rev.\  D {\bf 77}, 075010 (2008)
  [arXiv:0712.4019 [hep-ph]]; 
  J.~Garayoa and T.~Schwetz,
  JHEP {\bf 0803}, 009 (2008)
  [arXiv:0712.1453 [hep-ph]]; 
  M.~Kadastik, M.~Raidal and L.~Rebane,
  Phys.\ Rev.\  D {\bf 77}, 115023 (2008)
  [arXiv:0712.3912 [hep-ph]]; 
  P.~Fileviez Perez, T.~Han, G.~y.~Huang, T.~Li and K.~Wang,
  Phys.\ Rev.\  D {\bf 78}, 015018 (2008)
  [arXiv:0805.3536 [hep-ph]]; 
  F.~del Aguila and J.~A.~Aguilar-Saavedra,
  Nucl.\ Phys.\  B {\bf 813}, 22 (2009)
  [arXiv:0808.2468 [hep-ph]]; 
S.~T.~Petcov, H.~Sugiyama and Y.~Takanishi,
  Phys.\ Rev.\  D {\bf 80}, 015005 (2009)
  [arXiv:0904.0759 [hep-ph]]; 
  H.~Nishiura and T.~Fukuyama,
  Phys.\ Rev.\  D {\bf 80}, 017302 (2009)
  [arXiv:0905.3963 [hep-ph]].

\bibitem{han}
  T.~Han, B.~Mukhopadhyaya, Z.~Si and K.~Wang,
  Phys.\ Rev.\  D {\bf 76}, 075013 (2007)
  [arXiv:0706.0441 [hep-ph]]; 
 S.~Godfrey and K.~Moats,
  arXiv:1003.3033 [hep-ph]; 
see also K.~Huitu, J.~Maalampi, A.~Pietila and M.~Raidal,
  Nucl.\ Phys.\  B {\bf 487}, 27 (1997)
  [arXiv:hep-ph/9606311].


\bibitem{minimal}
  M.~Cirelli, N.~Fornengo and A.~Strumia,
  Nucl.\ Phys.\  B {\bf 753}, 178 (2006)
  [arXiv:hep-ph/0512090].




\end{thebibliography}
\end{document}